\documentclass[12pt]{article}

\usepackage{epsfig}

\usepackage{amssymb}

\def\be{\begin{equation}}
\def\ee{\end{equation}}
\def\ba{\begin{array}{c}}
\def\ea{\end{array}}

\newcommand{\bea}{\begin{eqnarray}}
\newcommand{\eea}{\end{eqnarray}}

\def\bp{\begin{proof}}
\def\ep{\end{proof}}

\newtheorem{thm}{Theorem}

\newenvironment{proof}{\noindent {\bf Proof.\ }}{\hfill$\square$\vspace{3mm}\endtrivlist}

\begin{document}


\begin{center}

{\Large \bf

Displaced harmonic oscillator
$V\sim \min
\,[(x+d)^2,(x-d)^2]$
as a benchmark double-well quantum model

  }

\vspace{9mm}

{Miloslav Znojil}

\vspace{9mm}

\vspace{0.2cm}

Department of Physics, Faculty of Science, University of Hradec
Kr\'{a}lov\'{e},

Rokitansk\'{e}ho 62, 50003 Hradec Kr\'{a}lov\'{e},
 Czech Republic

\vspace{0.2cm}

 and

\vspace{0.2cm}

The Czech Academy of Sciences, Nuclear Physics Institute,

 Hlavn\'{\i} 130,
250 68 \v{R}e\v{z}, Czech Republic

\vspace{0.2cm}

%
%
%
%

{e-mail: znojil@ujf.cas.cz}


\end{center}



\section*{Abstract}

For the displaced harmonic double-well oscillator
the existence of exact
polynomial bound states
at certain displacements $d\,$
is
revealed. The $N-$plets of these quasi-exactly solvable (QES)
states are
constructed in closed form.
For non-QES states, Schr\"{o}dinger equation
can still be considered ``non-polynomially exactly solvable'' (NES)
because the exact left and right parts of the
wave function
(proportional to confluent hypergeometric function)
just have to be matched
in the origin.

\subsection*{Keywords:}

.

displaced harmonic oscillators;

matching-method solutions;

quasi-exact and non-polynomial exact bound states;

double-well -- single-well transition;

\subsection*{PACS:}

.

 03.65.Ge  Solutions of wave equations: bound states

 02.30.Gp  Special functions

\newpage

\section{Introduction\label{Introduction}}

In the traditional realistic
applications of quantum mechanics
one picks up
a tentative one-dimensional
potential $V(x)$
in a way inspired, say, by the principle of correspondence.
Subsequently one
solves the
corresponding ``user-friendly''
(i.e., ordinary differential) bound-state
Schr\"{o}dinger equation
 \be
 -\, \frac{{\rm d}^2}{{\rm d} x^2}\, \psi_n(x)
 + V(x)\, \psi_n(x)= E_n\,
 \psi_n(x)\,,\ \ \ \ \psi_n(x) \in L^2(\mathbb{R})\,,
 \ \ \ \ \ n = 0, 1, \ldots\,
   \label{SEx}
  \ee
numerically, arriving at the acceptance
or rejection of the
model in dependence on the
agreement or disagreement
between the predictions and the measurements \cite{Fluegge}.

The task becomes simplified when the needs of the quantum phenomenology are met
by a function $V(x)$ which is analytic.
Then it makes sense to restrict attention to the potentials which are
exactly solvable (ES, \cite{Cooper}) or at least partially
exactly {\it alias\,} ``quasi-exactly'' solvable (QES, \cite{Ushveridze}).
Indeed, the
``redundant'' analyticity requirement may prove useful, e.g., when one
treats the elementary
Schr\"{o}dinger equation (\ref{SEx})
as a mere methodical laboratory needed,
say, during the study of supersymmetric systems \cite{Witten}.

Whenever a suitable
ES or QES model is not available,
people feel motivated to assign the
solvability status also
to some
non-analytic
models $V(x)$. Typically, these models
are
of the piecewise constant square-well type \cite{quesne}.
In applications,
their decisive formal merit is that
their bound-state wave functions $\psi_n(x)$ can be
defined, using the matching method, in terms of
elementary trigonometric functions.

In between the analytic and non-analytic extremes
one finds a few parity-symmetric models $V(x)=V(-x)$
in which the analyticity is only broken in the origin.
Under this assumption it was possible to
assign the QES solvability status
to the spatially symmetrized quartic anharmonic oscillator \cite{[296]}
or to the symmetrized sextic anharmonic oscillator \cite{[296b]}.

In the non-polynomial exact solvability (NES)
context
the non-QES exact spectra and bound states
of the parity-symmetric models
proved obtainable,
for a few parity-symmetric potentials,
via the $x=0$ matching of
their exact special-function left and right parts $\psi_n^{(\pm)}(x)$ of the
wave functions.
This enabled one to assign the NES status to the
symmetrized exponential potential $V(x)\sim \exp |x|$ (with $\psi_n^{(\pm)}(x)$
expressed in terms of
Bessel functions
\cite{[300],[24a]}),
to its asymptotically decreasing alternative
\cite{[302]} and/or to the
symmetrized Morse potential $V(x)$
(i.e., to the sum $V(x)$ of the two exponentials for which
$\psi_n^{(\pm)}(x)$
appeared proportional to a confluent hypergeometric function
\cite{[297],[24]}).

In our present paper we intend to complement these results
by showing that in some of the parity-symmetric quantum models
and, in particular, for
 \be
 V(x)=V_{[d]}^{{}}(x)= \left \{
 \begin{array}{ll}
  (x-d)^2,& x>0,\\
  (x+d)^2,&x<0
 \ea
 \right .
  \label{dolenhol}
 \ee
in Eq.~(\ref{SEx}), the QES and NES regimes can coexist.
One might also add that
disproportionately, the ES
quantum systems
with the single-well harmonic-oscillator potential
 $
 V(x)=V^{(HO)}(x)=x^2$ is discussed in virtually any textbook on
quantum mechanics (cf., e.g.,~\cite{Messiah}). In contrast,
the most straightforward twinned, $d-$displaced generalization (\ref{dolenhol})
of the harmonic-oscillator model with $d \in \mathbb{R} \setminus \{0\}\,$
is never mentioned. In what follows, we
intend to fill the gap.

\subsection{Motivation in physics}

In the context of physics the above-mentioned NES
constructions were motivated by the needs of molecular physics
so that the non-analyticity of $V(x)$ in the origin appeared inessential.
Nevertheless, even in this area of physics
the analyticity of
the benchmark models $V(x)$
re-emerged as essential,
mainly due to the related
possibility of a qualitative analysis of
{\em quantum\,}
dynamics using the
mathematical tools of the
{\em classical\,} theory of catastrophes \cite{Thom}.

The simplest forms of the theory
(viz., the so called fold and cusp catastrophes) proved
mostly needed \cite{dell2}.
Let us just remind the readers briefly that
in this language the equilibrium of
a classical dynamical system
is identified with a minimum
of a suitable ``Lyapunov'' potential $V(x)$
\cite{Zeeman}.
This enables one to classify
all of the qualitatively different patterns of
the bifurcation of the equilibria
in terms of the parameter-dependence of the minima.

By far the most interesting and useful example
of such a bifurcation pattern is called cusp
catastrophe. In this case
one conventionally chooses
 \be
 V(x)=V^{(cusp)}(x,a,b)=x^4+ax^2+bx\,.
 \label{moca}
 \ee
It is then easy to verify that such a Lyapunov function
possesses a single minimum at $a>0$ but two minima (i.e., a double-well shape)
at $a<0$ and at a not too large $b$. It is also worth noting that
in the $a-b$ plane the boundary of transition
between the two regimes is a spiked, cusp-shaped curve, indeed \cite{wiki}.
Thus, one can use the formalism for a qualitative description
of dynamics, say, of the Josephson junction between
the two Bose-Einstein condensates \cite{dell2b}.

After one turns attention to the
genuine quantum forms of dynamics,
the work with the classical $a-b$ plane geometry of model (\ref{moca})
ceases to be satisfactory.
One can expect that
the classical bifurcation phenomena become smeared out by the
quantization.
In particular, in the double-well case, due to the tunneling,
one would only have to move far away from the fine-tuned
cusp singularity
in order to be able
to distinguish, experimentally, between
the low-lying states which remain localized, in general,
either near the left minimum $R_-$ or near the right minimum
$R_+$ of $V^{(cusp)}(x,a,b)$.

In the latter, deep-double-well dynamical regime
the tunneling becomes suppressed since
the growth of the distance $R_+-R_-$
between the minima
becomes accompanied by their increasing separation mediated
by a growing and thickening barrier \cite{ArnoldI}.
The approximate calculation
of the bound states may be then facilitated
by the use of a fairly efficient
local harmonic-oscillator approximation,
 $$
 V^{(cusp)}(R_\pm + \xi,a,b)= V^{(cusp)}(R_\pm,a,b)+\omega^2_\pm\,\xi^2
 +{\cal O}(\xi^3)\,.
 $$
More or less equivalently, one may replace $V^{(cusp)}(x,a,b)$
by a global double-well harmonic oscillator,
 \be
 V(x,a,b)= \min
 \,[
 \omega_-^2\,(x-R_-)^2 +D_-\,,\,\omega_+^2\,(x-R_+)^2 +D_+
 ]\,.
 \label{ujfb}
 \ee
Whenever the difference
 $\,|\omega_--\omega_+|\,
 $
or
 $
 |V^{(cusp)}(R_-,a,b)-V^{(cusp)}(R_+,a,b)|\,$ becomes large,
the low-lying states remain localized
just near one of the minima.
The realization of a genuine quantum
effect of
bifurcation may only be expected to occur
when
$\omega_-^2\approx \omega_+^2$ and  $D_-
\approx D_+$.
Thus, we arrive at the
spatially symmetric
toy-model potential
 \be
 V^{}(x,d,\omega)=\omega^2\,\min
 \,[
 (x+d)^2,(x-d)^2
 ]\,
 \label{ujf}
 \ee
in which we may scale out $\omega \to 1$ and obtain the
ultimate double-well version of the one-parametric
potential (\ref{dolenhol}) of our present interest.


\begin{figure}[h]                    
\begin{center}                         
\epsfig{file=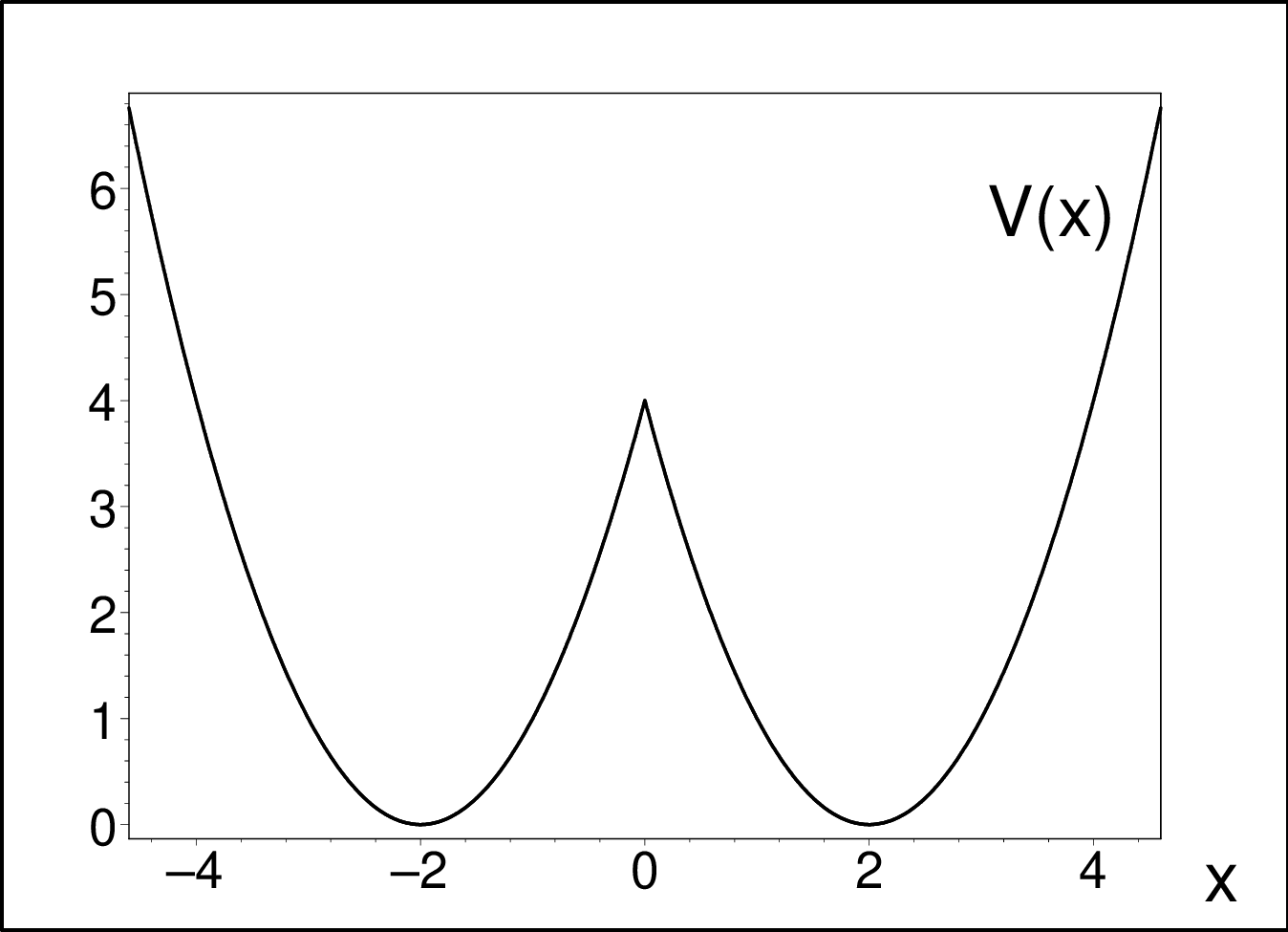,angle=270,width=0.30\textwidth}
\end{center}    
\vspace{2mm} \caption{Double-well shape of potential
(\ref{dolenhol}) at positive $d=2$.
 \label{svizt}
 }
\end{figure}

It is worth adding that
our asymptotically quadratic one-dimensional potential (\ref{dolenhol}), viz.,
$V_{}(x,d,1)=x^2-2d\,|x|+d^2$ with $d \neq 0\,$
represents in fact just a single-well interaction at a negative
displacement $d<0$, and it only acquires the more interesting
double-well shape at a positive $d>0$
(see its sample in Figure \ref{svizt} where $d=2$).

\subsection{Motivation in mathematics}


At the not too large displacements $d$,
the elementary one-dimensional potential (\ref{dolenhol})
can be peceived as an example of a smooth
perturbation of an ubiquitous harmonic oscillator. Still,
the non-analyticity of $V(x)$ in the origin
looks ``suspicious''. For this reason, in spite of the relative
boundedness of the perturbation, the model seems never recalled in
the context of perturbation theory. It sounds almost like a paradox
that in the textbooks on quantum mechanics the role of the most
popular illustrative example of the perturbation of harmonic
oscillator is almost invariably played by the (nicely analytic)
quartic anharmonicity $\sim x^4$ yielding the {\em vanishing} radius
of convergence of the resulting Rayleigh-Schr\"{o}dinger
perturbation series \cite{zero}.

In our recent comment on quartic anharmonicities \cite{symqua} we
pointed out that the not too rational (perhaps, purely emotionally
motivated) insistence of the major part of the physics community on
the strict analyticity of the one-dimensional phenomenological
potentials $V(x)$ did also cause problems in the monograph
\cite{Ushveridze}. Indeed, the author of this very nice review of
the so called quasi-exactly solvable (QES) Schr\"{o}dinger equations
(offering also an extensive list of further relevant references) did
not imagine that besides the best known sextic-polynomial QES
potential $V(x)$, the QES status can be also assigned,
in a way explained in \cite{symqua}, to its
quartic-polynomial predecessor, provided only that we admit its
non-analyticity at $x=0$.

Recently we returned to the QES problem and imagined that even the
quartic polynomials $V(x)=A|x|+Bx^2+C|x|^3+x^4$ which were studied
and assigned the QES status in the latter reference need not still
represent the ``first nontrivial'' case. Our subsequent study of the
problem resulted in the conclusion that the role of
the simplest nontrivial QES quantum model can in fact be played by
the spiked harmonic oscillator (\ref{dolenhol}).

A concise constructive demonstration of this assertion is to be
presented in the following section.

\section{Polynomial solvability}

Our model (\ref{dolenhol}) is important as offering a
{\em qualitative} description of the phenomena connected with the
changes of the topology of the forces as characterized by the so called
pitchfork bifurcation of equilibria
in classical
mechanics
\cite{Zeeman}. In quantum context
formula~(\ref{dolenhol}) interpolates between the exactly solvable
limit $d \to +\infty$ (yielding the doubly degenerate spectrum of
energies $E_n^{\pm}[\omega,\infty] = (2n+1)\omega$, $n = 0,1,
\ldots$, where the superscript marks the parity of the related wave
functions) and {\em another}  exactly solvable limit $d \to 0$
providing {\em the same, but non-degenerate}, spectrum of energies
$E_n[\omega,0] = (2n+1)\omega$, $n = 0,1, \ldots\,$ with parities
$(-1)^n$.
In between the wave functions
have to be derived as satisfying the physical boundary
conditions in infinity as well as in the origin.

\subsection{The matching of states $\psi(x)$ in the origin}

The phenomenological appeal of the genuine quantum bifurcations often
forces us to turn attention to
approximate methods. {\it Pars pro toto\,} let us mention
our recent studies \cite{ArnoldI,ArnoldII} of the phenomenon of tunneling
in which an analytic (in fact, polynomial) potential $V(x)$
possessing
an $N-$plet of the local minima $V(R_j)$, $j = 1, 2, \ldots, N$
has been replaced,
in the low-lying spectrum regime,
by a locally adapted harmonic-oscillator well as sampled,
at $N=2$, by Eq.~(\ref{ujfb}).
Beyond the $N=2$ harmonic-oscillator approximation
as well as beyond $N=2$,
the construction of the
wave-functions
using the matching of the logarithmic derivatives at
the $(N-1)-$plet of non-analyticities
becomes purely numerical.

In this sense our present $N=2\,$ toy model
is exceptional.
At any (i.e., not necessarily just at bound-state) energy $E$
its exact and correct wave functions
$\psi^{(\pm)}(x,E)$ (which vanish
in infinity and could be called, therefore, the Jost solutions)
are proportional to certain
special functions which can be considered
known (see their explicit form given a bit later below).
Thus, one only has to
match the logarithmic derivatives
of the resulting left branch $\psi^{(-)}(x,E)$
and right branch $\psi^{(+)}(x,E)$
of $\psi(x)$
in the origin,
 \be
 \lim_{x \to 0}\psi^{(+)}(x,E)=\lim_{x \to 0}\psi^{(-)}(x,E)\,,\ \ \  \ \ \
 \lim_{x \to 0}(\psi^{(+)})'(x,E)=\lim_{x \to 0}(\psi^{(-)})'(x,E)\,.
 \ee
As long as the ultimate, properly matched
wave function has to be of an even or odd parity,
we may write $\psi(x,E) =\psi(x,E,p) = (-1)^p\,\psi(-x,E,p)$
with $p = 0$ or $p=1$, respectively. Thus, under certain non-singularity
conditions in the origin
we may
introduce an arbitrary regular {\it ad hoc\,} normalization
 \be
 \lim_{x \to 0}\psi^{(+)}(x,E,0)={\cal N}_0\neq 0\,,\ \ \ \ \ \
 \lim_{x \to 0}(\psi_n^{(+)})'(x,E,1)={\cal N}_1\neq 0
 \,
 \ee
and arrive at the
two alternative ``shooting method'' matching rules
 \be
 \lim_{x \to 0}(\psi^{(+)})'(x,E,0)=0\,,\ \ \ \ \ \
 \lim_{x \to 0}\psi^{(+)}(x,E,1)=0
 \,.
 \label{MR}
 \ee
In this spirit also the above-cited samples of a few other NES
constructions of bound states can be perceived as
the implementations of the same matching rules {\it alias\,}
transcendental equations (\ref{MR}). In all of these cases,
these equations
determine the whole spectrum of the bound state energies $E=E_n$
in implicit manner.

\subsection{The simplest polynomial solution\label{uQESsec}}


Our interest in one-dimensional potential (\ref{dolenhol}) was born
when we noticed that it is quasi-exactly solvable (QES). By
definition \cite{Ushveridze} this means that it admits the existence
of exact elementary wave functions at certain parameters and
energies. One can easily verify, by insertion, that at a
special shift $d=d^{(QES)}=-1$ the nodeless function
 \be
\psi_0^{(QES)} \left( x \right) = \left \{
 \begin{array}{ll}
 \left(1+ x\right) {e^{- \left(
 x^2/2 +x\right) }}, \ \  \ x>0,\\
 \left( 1-x\right) {e^{- \left(
 x^2/2 -x\right) }}, \ \  \ x<0
 \ea
 \right .\,
 \label{waqvef}
 \ee
belongs to $L^2(\mathbb{R})$, obeys the standard matching conditions (i.e.,
has a continuous logarithmic derivative in the origin) and satisfies
our Schr\"{o}dinger Eq.~(\ref{SEx}) + (\ref{dolenhol}) at
$E=E_0^{(QES)}=3$ exactly. Thus, this expression represents the even-parity
ground-state wave function of the model.


A non-constructive proof of the QES property of our model is
elementary. In the single-well scenario we choose the special value
of $d=d^{(QES)}=-1$ and inserted the nodeless function
(\ref{waqvef}) (which belongs to $L^2(\mathbb{R})$) in
Schr\"{o}dinger equation
 \be
 -\, \frac{{\rm d}^2}{{\rm d} x^2} \psi(x)
 + V(x) \psi(x)= E\,
 \psi(x)\,.
   \label{reSEx}
  \ee
Clearly, function (\ref{waqvef}) obeys the standard matching
conditions  in the origin (i.e., it has a continuous logarithmic
derivative there) and it solves our bound-state
problem~(\ref{dolenhol}) + (\ref{reSEx})  at $E=E_0^{(QES)}=3$.
Thus, QES expression (\ref{waqvef}) represents the {\em exact}
even-parity ground-state wave function. In Fig.~\ref{qesvizt} our
QES choice of parameters has been used for illustration purposes.
The similar elementary QES feature also occurs after one moves to
the double-well dynamical regime.


%
\begin{figure}[h]                    
\begin{center}                         
\epsfig{file=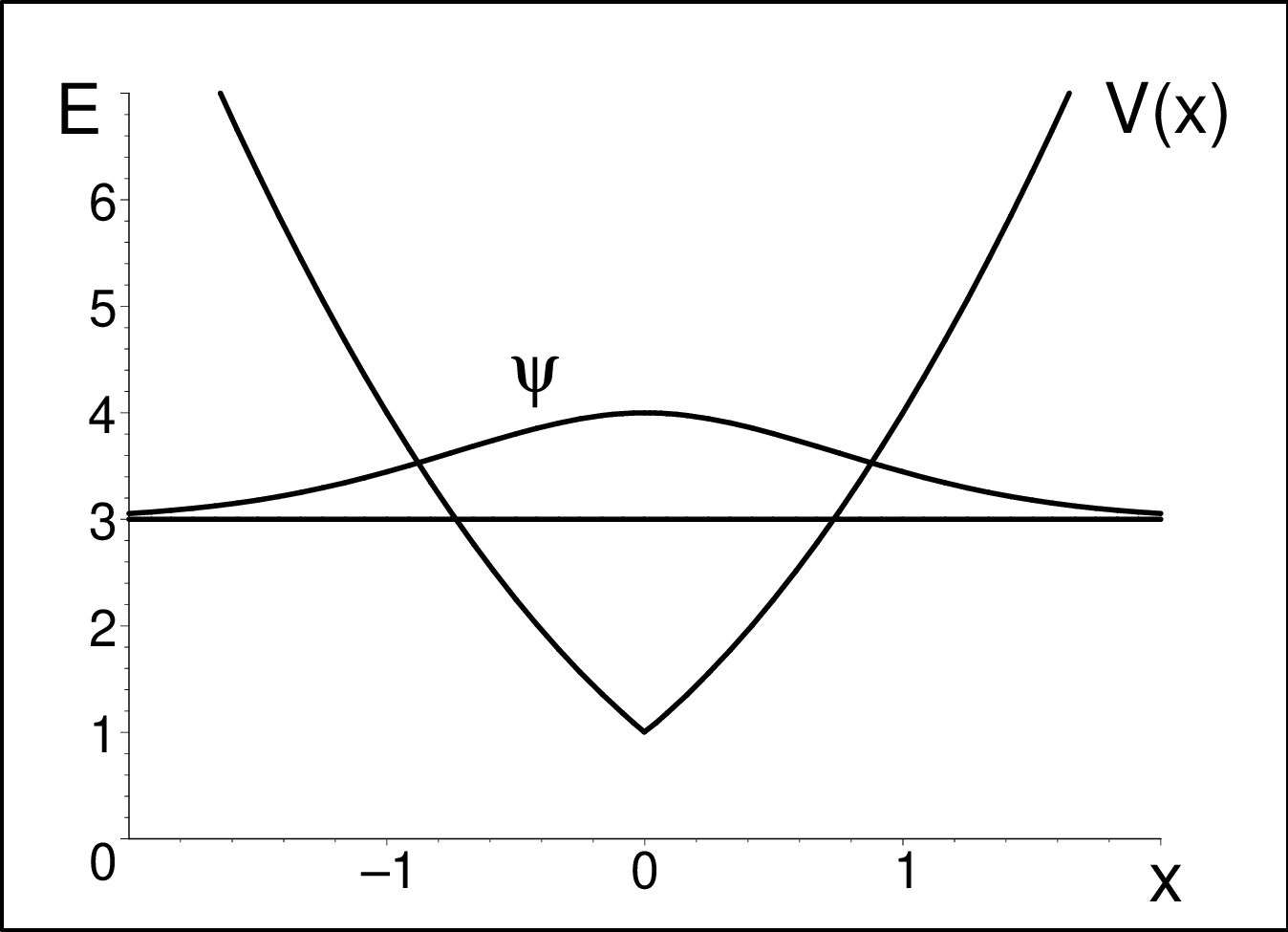,angle=270,width=0.30\textwidth}
\end{center}    
\vspace{2mm} \caption{Ground-state energy and wave function in the
single-well QES version of potential (\ref{dolenhol}) at
$d=-\mu=-1$.
 \label{qesvizt}
 }
\end{figure}

Let us add that in the standard literature the QES property has only
been assigned to the 1D potentials which were manifestly analytic
along the whole real line \cite{Singh,Turbiner,Fring}. The present,
non-analytic generalization of the QES property has only been
proposed in Ref.~\cite{symqua}. The
non-analytic QES construction has been preformed there for the
phenomenologically important class of quartic anharmonic
oscillators. Our present real QES wave function (\ref{waqvef}) may
be also recalled as providing a closed-form illustration of the fact
that besides the potential, also our wave functions are non-analytic
at $x=0$ since
 \be
 \frac{d^3}{dx^3}\psi_0^{(QES)} \left( 0^+ \right)=2 \neq
 \frac{d^3}{dx^3}\psi_0^{(QES)} \left( 0^- \right)=-2\,.
 \ee
We are persuaded that the descriptive merits of the similar
generalized QES constructions overweight their not too essential
feature of non-analyticity at $x=0$. In this sense the existing
literature would certainly deserve to be completed.

\subsection{Systematic approach}

After one inserts potential (\ref{dolenhol}) in Schr\"{o}dinger
Eq.~(\ref{reSEx}) it is possible to assume the existence of the
even-parity and/or odd-parity QES bound-state solutions in the most
general form of the standard normalizable polynomial ansatz such
that, on the positive half-axis,
 \be
\psi^{(QES)} \left( x \right) =\psi \left( x \right) =
  e^{-
 x^2/2 +d x}\,
 \times
  \sum_{k=0}^N
  \,a_kx^k\,, \ \ \ \ \ x \geq  0\,,\ \ \ \ a_N \neq 0\,.
 \label{waqvefgen}
 \ee
We may accept the normalization convention
 \be
 \left \{
 \begin{array}{cc}
 \psi \left( 0 \right)=1\,,&
 \psi' \left( 0 \right)=0\,,\\
 \psi \left( 0 \right)=0\,,&
 \psi' \left( 0 \right)=1\,,
 \ea
 \right .
 \ \ \
 {\rm i.e.,}
 \ \ \
 \left \{
 \begin{array}{cccc}
 a_0=1\,,&
 a_1=-d\,,&& {\rm parity = even}\,,\\
 a_0=0\,,&
 a_1=1\,,&a_2=-d\,,&{\rm parity = odd}\,.
 \ea
 \right .
 \label{bc}
 \ee
With $a_{N+1}=a_{N+2} = \ldots = 0$  and after appropriate
insertions and elementary algebra the QES solvability of the model
appears equivalent to the validity of the set of linear recurrences
 \be
 (E-1-2n)\,a_n+ 2d(n+1)\,a_{n+1}+(n+1)(n+2) \,a_{n+2}=0\,,
 \ \ \ \ n = 0, 1, \ldots , N\,.
 \label{rekure}
 \ee
From the last item with $n=N$ we get the QES value of energy
$E=2N+1\ $ so that we are left with an $N-$plet of relations between
the $N-1$ unknown coefficients $a_2$, $a_3$, \ldots , $a_N\ $ and an
unknown, QES-compatible value (or rather a multiplet of values) of
the shift $d=d^{(QES)}$.

Starting from the choice of $N=0$ and $E=1$ we find that the odd
solution cannot exist while the well known even-parity solution only
exists at $d=0$. In the next step with $N=1$ and $E=3$ the single
constraint $2\,a_0+2\,d\,a_1=0$ only admits the pure $d=0$ harmonic
oscillator in the odd-parity case while the first nontrivial QES
solutions with $d=\pm 1$ emerges in the even-parity scenario.

Once we  skip the harmonic-oscillator solutions and demand that $d
\neq 0$  we may set $N=2$ and $E=5$ in Eq.~(\ref{rekure}). Then we
have to satisfy the set of two relations
 \be
 4a_0+2da_1+2a_2=0\,,
 \ \ \ \ \
 2a_1+4da_2=0\,.
 \ee
This leads to the two easy odd-parity solutions with $d=d_\pm = \pm
1/\sqrt{2}$ and $a_2=-d$. They become accompanied by the equally
easy even-parity solutions with $a_2=1/2$ and $d=\pm \sqrt{5/2}$.

Next we choose $N=3$ and find that the odd-parity construction
degenerates to the triplet of relations
  $$
   2+2\,d\,{\it a_2}+3\,{\it a_3}=0,
   \ \ \
  \,d+{\it a_2}=0,
  \ \ \ \ \,{\it a_2}+3\,d\,{\it a_3}=0
  $$
with solution $a_3=1/3$, $d=\pm \sqrt{3/2}$ and $a_2=-d$. In
parallel, the even-parity conditions
 $$
 -2\,d\,+2\,d\,{\it a_2}+3\,{\it a_3}=0,
 \ \ \ 3-{d}^{2}+{\it
a_2}=0, \ \ \ {\it a_2}+3\,d\,{\it a_3}=0
 $$
lead to the slightly less trivial quadruplet of eligible QES shifts
 \be
 d_{\pm,\pm}=\pm \sqrt{\frac{9\pm \sqrt{57}}{4}}\,.
 \ee
Each one of them defines the related two coefficients
 \be
 a_2=\frac{2d^2}{2d^2-1}\,,\ \ \ \
 a_3=\frac{2d}{6d^2-3}\,.
 \ee

\subsection{General case}

\begin{table}[h]
\caption{QES even-parity values of shifts $d$ are defined,
implicitly, as zeros of polynomials $P^{(N)}(d)$. A sample of the
Gr\"{o}bnerian elimination is added.}
 \label{Pwe1}
\centering
\begin{tabular}{||c||c|c||}
\hline \hline
  \multicolumn{1}{||c||}{}& polynomial &\multicolumn{1}{|c||}{\rm elimination}
 \\
  \multicolumn{1}{||c||}{$N$}& $P^{(N)}(d)$& of $a_N$
 \\
 \hline
 2 &$ -5+2\,d^2$& $ -1+2\,a_2=0$\\
  3 &$ 3-9\,d^2+2\,d^4$& $ -8\,d+2\,d^3+3\,a_3=0$\\
  4 &$27-28\,d^2+4\,d^4$& $ 11-2\,d^2+12\,a_4=0$\\
    5 &$-15+75\,d^2-40\,d^4+4\,d^6$& $ 29\,d-19\,d^3+2\,d^5+15\,a_5=0$\\
 \vdots &$ \ldots$& $\ldots$\\
 \hline \hline
\end{tabular}
\end{table}

 \noindent
At any  $N$ we may study the QES sets of coupled nonlinear algebraic
equations using the computer-assisted elimination technique of
Gr\"{o}bner bases. In the even-parity case this algorithm generates
certain polynomials $P^{(N)}(d)$ which specify the QES-compatible
shifts $d$ as their zeros (see Table~\ref{Pwe1}). The odd-parity QES
solutions appear determined, similarly, via the values of $a=a_2=-d$
which coincide with the real zeros of other polynomials $Q^{(N)}(a)$
(see Table~\ref{Qwe1}).

\begin{table}[h]
\caption{QES values $a=a_2=-d$ determined as zeros of polynomials
$Q^{(N)}(a)$ (odd-parity case).}
 \label{Qwe1}
\centering
\begin{tabular}{||c||c|c||}
\hline \hline
  \multicolumn{1}{||c||}{}& polynomial &\multicolumn{1}{|c||}{\rm elimination}
 \\
  \multicolumn{1}{||c||}{$N$}& $Q^{(N)}(a)$& of $a_N$
 \\
 \hline
 2 &$ -1+2\,a^2$& $ -a+a_2=0$\\
  3 &$ -3+2\,a^2$& $ -1+3\,a_3=0$\\
  4 &$ 3-12\,a^2+4\,a^4$& $ 5\,a-2\,a^3+6\,a_4=0$\\
 5 &$ 15-20\,a^2+4\,a^4$& $ 7-2\,a^2+30\,a_5=0$\\
 \vdots &$ \ldots$& $\ldots$\\
 \hline \hline
\end{tabular}
\end{table}


The most compact representation of the above set of results may be
obtained when one imagines that the even- and/or odd-parity boundary
condition (\ref{bc}) at $x=0$ may be reinterpreted as the respective
initial, additional (i.e., $n=-1$) condition for recurrences
(\ref{rekure}). This enables us to solve these recurrences in terms
of compact formulae.

\begin{thm}
In the even-parity QES case the general closed-form solution of
recurrences (\ref{rekure}) reads
 \be
 a_k=a_k^{(N)}=\frac{(-1)^k}{k!(k-1)!}
 \det {\cal P}^{(N,k)}(d)\,,\ \ \ k = (1, 2,)\, 3,4,\ldots\,.
 \label{rekbecplus}
 \ee
Its $d-$dependence is encoded in the tridiagonal $k$ by $k$ matrices
 \be
 {\cal P}^{(N,k)}(d)=
 \left (
 \begin{array}{llllll}
 d&1&0&0&\ldots&0\\
 2N&2d&2&0&\ddots&\vdots\\
 0&2N-2&4d&6&\ddots&0\\
 0&\ddots&\ddots&\ddots&\ddots&0\\
 \vdots&\ddots&0&2(N-k+3)&2(k-2)d&(k-1)(k-2)\\
 0&\ldots&0&0&2(N-k+2)&2(k-1)d\\
 \ea
 \right )\,.
 \label{tamatsu}
 \ee
These formulae must be complemented by the constraint
$a_{N+1}^{(N)}=0$, i.e., by the polynomial algebraic equation
 \be
 \det {\cal P}^{(N,N+1)}(d)=0
 \ee
which represents an implicit definition of all of the admissible
values of the specific, QES-compatible shift parameters
$d=d^{(QES)}$.
\end{thm}
 \bp
One may proceed by mathematical induction. It is only necessary to
keep in mind that in matrix~(\ref{tamatsu}) the first row reflecting
the initial condition looks (and is) anomalous. That's why the form
and validity of the $k=1$ solution $a_1=-d$ and also of the $k=2$
solution $a_2=d^2-N$ are to be verified separately.
 \ep

\begin{thm}
In the odd-parity QES case the closed-form solution of recurrences
(\ref{rekure}) reads
 \be
 a_{k+1}=a_{k+1}^{(N)} =\frac{(-1)^{k}}{(k+1)!k!}
 \det {\cal Q}^{(N,k)}(d)\,,\ \ \ k = 1, 2, \ldots\,
 \label{rekbecminus}
 \ee
with the $d-$dependence encoded in the tridiagonal $k$ by $k$
matrices
 \be
 {\cal Q}^{(N,k)}(d)=
 \left (
 \begin{array}{lllll}
  2d&2&0&\ldots&0\\
 2N-2&4d&6&\ddots&\vdots\\
 0&2N-4&\ddots&\ddots&0\\
 \vdots &\ddots&\ddots&2(k-1)d&k(k-1)\\
 0&\ldots&0&2(N-k+1)&2kd\\
 \ea
 \right )\,.
 \label{furu}
 \ee
This  must be complemented by the algebraic polynomial-equation
definition $a_{N+1}^{(N)}=0$, i.e.,
 \be
 \det {\cal Q}^{(N,N)}(d)=0
 \ee
yielding all of the admissible values of the specific,
QES-compatible shift parameters $d=d^{(QES)}$.
\end{thm}
 \bp
Once one keeps in mind that in the odd-parity cases we have $a_2=-d$
at all $N$ and $k$, the proof remains elementary and proceeds, along
similar lines as above, by mathematical induction.
 \ep

\section{Special-function solvability}

In the broader context of quantum physics the amount of attention
which is currently paid to polynomially solvable
(traditionally called ``exactly  solvable'', ES) Schr\"{o}dinger
equations seems disproportionate. Especially in comparison with the
fact that the attention virtually disappears
in the NES cases, i.e.,
immediately behind the
latter ES area. The paradox is striking because
the sophisticated commercial software (like
MAPLE or MATHEMATICA, etc)
as well as the freely available NIST review \cite{nist}
enable the physicists to work with
elementary functions as easily as with the special functions.

Within the NES approach one can decide to
treat the terminating and non-terminating confluent
hypergeometric series on equal footing. As an immediate consequence of such an
amendment of the paradigm
one obtains an extension of the class of the user-friendly
interactions.
In our present paper we accept the
idea and intend to describe
one of its most natural implementations in the one-dimensional
quantum-dynamical scenario.

In this context one can feel inspired by the classical Thom's
concepts of stability and of its loss are
closely related to the algebraic
geometry of the minima of certain {\em
analytic} Lyuapunov's functions $V(x)$ \cite{Arnold}. For this reason the
classical dynamics near a cusp-related bifurcation (i.e., near an
``abrupt'' transition between one and two stable minima of
$V(x)$) is traditionally described via a
representative {\em analytic}, anharmonic-oscillator (AHO) polynomial
(\ref{moca}).
It is widely believed that
the polynomial Lyapunov's functions $V(x)$ are best suited
for the study of
the Thom's classical catastrophes
and that even the constructions of their quantum analogues
should not break the correspondence with these polynomials.
Thus, even though our present model $V^{{}}_{[d]}(x)$ could have
also pretended to play the role of a cusp-related Lyapunov function
(especially in the simplest nontrivial bifurcation scenario with
$b=0$ in AHO model), it seemed disqualified by its
non-analyticity in the origin.

\subsection{Single-well dynamical regime with $d<0$\label{2.0}}

Very recently we imagined that any quantization of a classical catastrophe
must replace the usual schematic time-dependent trajectories
$x=x(t)$ in a classical space of states by certain selfadjoint
operators $\hat{x}(t)$ in a suitable infinite-dimensional physical
Hilbert space ${\cal H}^{(P)}$ \cite{katast}. Thus, after quantization the
concepts like bifurcation (etc) must necessarily be transferred from
the domain of algebraic geometry to the area of functional analysis
and spectral theory. Naturally, this implies that the
non-analyticity of $V^{{}}_{[d]}(x)$ becomes inessential while its
solvability may start playing the role of a significant merit in
comparison with a purely numerical nature of AHO Schr\"{o}dinger
equations \cite{AHOs}. In other words,
the twinned harmonic oscillator
(\ref{dolenhol}) with a variable real displacement $d$ and with two minima
iff $d>0$ seems suitable
and acceptable as a benchmark quantum model with catastrophic features.

Besides the serendipitous and rather formal QES property
as discussed above, a turn of
attention to the fully general
twinned harmonic-oscillator model (\ref{dolenhol})
is recommendable from phenomenological perspective. Another,
independent reason would be again mathematical. After an
elementary change of variables one reveals that Eq.~(\ref{SEx})
coincides with the two branches of the confluent hypergeometric
equation which are matched together in the origin. In this sense our
model (\ref{SEx}) + (\ref{dolenhol}) resembles square wells and
remains solvable via the matching of logarithmic derivatives of $\psi(x)$ at
$x=0$.
In the literature the conceptual appeal of the complete and, in
principle, arbitrarily precise solvability based on the use of
confluent hypergeometric special functions may be found advocated
by multiple authors
\cite{Fluegge}. In what follows we will defend and accept such a NES
point of view, demonstrating its applicability to the one-parametric
family of potentials (\ref{dolenhol}) in both of their one-centered
and
two-centered realizations.

Once we recall  the left-right symmetry
(i.e., parity-symmetry, ${\cal P}-$symmetry) of our Hamiltonian, we
may temporarily restrict our attention to the differential
Eq.~(\ref{SEx}) on one of the half-lines (i.e., say, with $x \in
(0,\infty)$). Then we may change variables as follows,
 $$
 x-d = \sqrt{z}  \in (-d,\infty) \,,\ \ \ \ \psi(x)=\exp(-z/2)\,w(z)\,.
 $$
Whenever we choose a negative $d=-\mu^2<0$, this converts
Eq.~(\ref{SEx}) into Kummer's equation
 \be
 z\frac{{d}^{2}w(z)}{{dz}^{2}}+(b-z)\frac{dw(z)}{dz}-aw(z)=0\,,
 \ \ \ \ \ z \in (\mu^4,\infty)
 \label{Kum}
 \ee
(cf. \cite{nist}, \S 13.2(i)) with parameters $a=(1-E)/4$ and
$b=1/2$.
As long as $z\geq \mu^2>0$ we may ignore the branch-point singularity of
Eq.~(\ref{Kum}) at $z=0$
and we may write down the general solution
as a superposition
of two independent confluent
hypergeometric series, viz.,
 $$
 w_{(even)}(z)=\mathop{M\/}\nolimits\! (a,b,z )=\ _1\!{F}_1(a,b,z)=
 $$
 \be
 =\sum_{s=0}^{\infty}\frac{\left(a%
 \right)_{s}}{\left(b\right)_{s}s!}z^{s}
 =1+\frac{a}{b}z+\frac{a(a+1)}{b(b+1)2!}%
 z^{2}+\cdots
 \ee
 and
 \be
 w_{(odd)}(z)= z^{1-b}\mathop{M\/}\nolimits\!  (a-b+1,2-b,z )\,.
 \label{defo}
 \ee
Subsequently, we have to satisfy the asymptotic physical boundary
condition. This forces us to combine the two independent solutions
$w_{(even)}(z)$ and $w_{(odd)}(z)$ of Eq.~(\ref{Kum}) in the
following unique manner,
 $$
 w_{(physical)}(z)=\mathop{U\/}
 \nolimits\!\left(a,b,z\right)
 =\frac{\mathop{\Gamma\/}\nolimits\!%
 \left(1-b\right)}{\mathop{\Gamma\/}\nolimits\!\left(a-b+1\right)}\mathop{M\/}%
 \nolimits\!\left(a,b,z\right)+
 $$
 \be
 +\frac{\mathop{\Gamma\/}\nolimits\!\left(b-1%
 \right)}{\mathop{\Gamma\/}\nolimits\!\left(a\right)}z^{1-b}\mathop{M\/}%
 \nolimits\!\left(a-b+1,2-b,z\right)\,.
 \label{asycorr}
 \ee
The choice of this function with the power-law asymptotical behavior
$\mathop{U\/}\nolimits\!\left(a,b,z\right) \sim z^{-a}$ valid in
the complex plane of $z=x+\mu^2$ with the left cut from $z=0$ to $-\infty$ yields the
correct, asymptotically vanishing wave functions
at positive $x\geq 0$ in closed form,
 $$
 \psi(x)=\psi(x,E)=
 $$
 \be
 =\exp(-(x+\mu^2)^2/2)\,
 \mathop{U\/}\nolimits\!\left((1-E)/4,1/2,(x+\mu^2)^2\right)
 \,.
 \ee
The formula remains manifestly dependent on the value of energy $E$.
After the matching of the left and right branches of our analytic
solution in the origin, the spectrum of bound states will be
obtained as split in the even-parity and the odd-parity states.
Their energies $E_n^{(\pm)}$ will coincide with the
roots of the respective transcendental secular equations, viz.,
 \be
 \psi(0,E^{(+)})=1\, \ \
 \psi'(0,E^{(+)})=0
 \label{secul}
 \ee
or
 \be
  \psi(0,E^{(-)})=0
  \ \   \psi'(0,E^{(-)})=1
 \,.
 \label{secula}
 \ee
Here the prime marks the derivative with respect to the first argument.


The numerical performance of the determination of the energies based
on the latter equations is entirely routine and reflects our
expectations that with the growth of the negative shift $\mu^2=-d$
the narrowing of the potential well will push the energy levels
upwards. Thus, the standard HO ground state with the energy equal to
one will be pushed up to   $E_0=E_0(-\mu^2)$ with $E_0(-0.25) \approx
1.3349$ and $E_0(-1)= 3$ (this is quasi-exact state), etc. The
similar accelerated growth moves the second excited state from
initial   $E_2(0)= 5$ more quickly to   $E_2(-1)\in (8.658,8.659)$,
etc. The bracketing property of the latter type of results is highly
welcome in computations of course.

%
%
%
%

\subsection{Double-well dynamical regime with $d>0$\label{2.1}}

In the double-well regime with positive $d=\nu^2>0$ the presence of
the square-root branch-point of Eq.~(\ref{Kum}) at $x=\nu^2$, i.e.,
{\em inside} the interval of physical coordinates would make the
direct use of the asymptotically correct closed-form solutions
(\ref{asycorr}) slightly uncomfortable. The key new technicality is
that one must keep the choice of the proper branch of the coordinate
function $x=x(z)= d \pm \sqrt{z}$ under explicit control.

For this reason we recommend the use of an alternative bound-state
construction in which one recalls the ${\cal P}-$symmetry of our
Schr\"{o}dinger Eq.~(\ref{SEx}) and employs a variant of the so
called shooting method. In the language of mathematics this means
that one simply separates the spatially even wave functions (for
which we set $\psi(0)=1$ and $\psi'(0)=0$ as an initial boundary
condition) from the spatially odd solutions (for which we have
$\psi(0)=0$ and $\psi'(0)=1$) -- this is to be compared with
Eqs.~(\ref{secul}) and (\ref{secula}), respectively.
In each of these two subcases one
arrives again at the Kummer's Eq.~(\ref{Kum}) which is now defined
in an extended and partially doubly covered interval of $z \in
(0,\infty)$ and $z \in (0,\nu^4)$.

In contrast to preceding subsection we will now solve the
initial-value {\it alias} Cauchy problem knowing that we know its
general solution in confluent-hypergeometric form
 \be
 \psi(x)=\psi(x,E)=C_1\, w_{(even)}[z(x)] + C_2\, w_{(odd)}[z(x)]
 \,.
 \label{olurso}
 \ee
In the next stage of development we fix the ratio of the
coefficients $C_j$ via the above-mentioned initial conditions
imposed upon the both-sided limits of $\psi(x)$ and
$\psi'(x)$ at $x=0$ (i.e., at $z = \nu^4$ when taken below its
square-root branch). The subsequent necessary control of the choice
of the branch of function  (\ref{defo}) (i.e., of the
analytic-function component $w_{(odd)}[z(x)]$ of our solution
(\ref{olurso}) at $z=0$) is not difficult because we have
$b=1/2$ (i.e., a square-root branch) in definition~(\ref{defo}).

In the last step we may finally recall the standard oscillation
theorems \cite{Ince} which

\begin{itemize}

\item
relate the quantum number $n=0,1,\ldots \,$ of a bound-state energy
$E_n$ to the number $Z_n=n$ of the finite nodal zeros $x_j\in
(-\infty,\infty)$ of $\psi_n(x)$;

\item
enable us to perturb $E_n \to E^{(+)}_n=E_n + \delta^2$, localize
the related maximal real (and, say, positive) finite nodal zero
$x_{(max)}^{(+)}>0$ and find the one-to-one correspondence between
the exact-bound-state-energy limits of $\delta\to 0$ and of
$x_{(max)}^{(+)} \to \infty$.

\end{itemize}

 \noindent
The latter correspondence enables us to localize, numerically, the
exact energy, with arbitrary preselected precision, via a
trial-end-error choice of $E^{(\pm)}_n$, to be iteratively amended
by a tentative, smaller and smaller decrease (whenever the maximal
real node $x_{(max)}^{(+)}$ keeps growing) or increase (whenever the
maximal node happens to leave the real line and become complex).

%

\begin{figure}[h]                    
\begin{center}                         
\epsfig{file=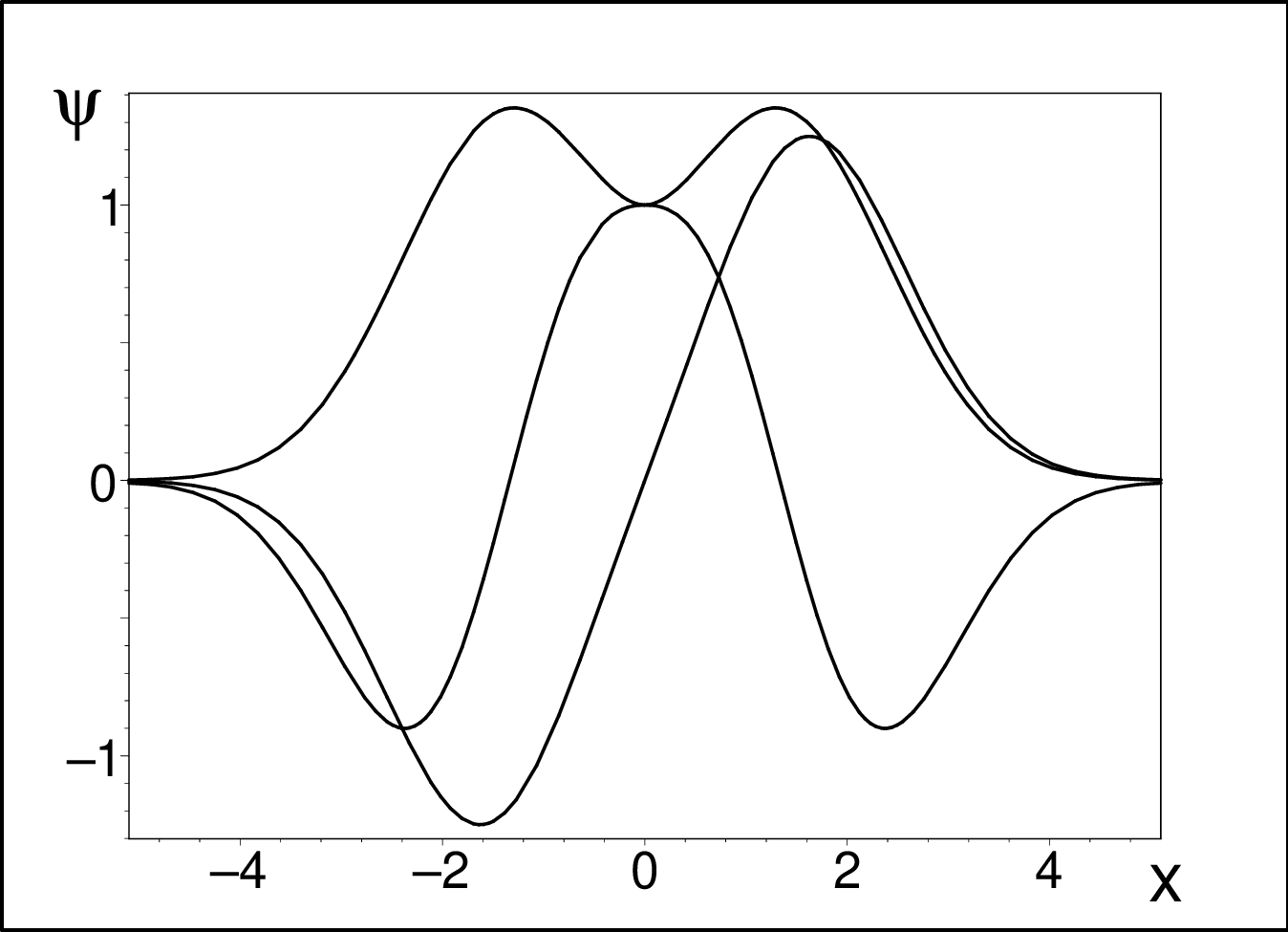,angle=270,width=0.30\textwidth}
\end{center}    
\vspace{2mm} \caption{The first three double-well wave functions at
$d=3/2$.
 \label{ffit}
 }
\end{figure}

\section{The displacement-dependence of the spectrum}

The numerical results of the above-outlined algorithm
are sampled in Figure \ref{ffit} and by Tables \ref{dowe} -
\ref{dowe2}. The key advantage of such a special-function-based
version of the well known shooting algorithm is that in our model
the trial-and-error functions $\psi(x)$ are known in the
exact (albeit rather clumsy, computer-stored) special-function
confluent-hypergeometric
form (which is closed and analytic but, unfortunately,
still too long for a printed display).
This enables
us to keep the numerical errors under active control and to
construct, systematically, the pairs of the lower and upper
estimates of the bound-state energies.

%
%
%
%


\begin{table}[h]
\caption{The shift-dependence of the ground-state energy.}
 \label{dowe}
\centering
\begin{tabular}{||c||c|c||}
\hline \hline
  \multicolumn{1}{||c||}{\rm {\rm shift}}&
 \multicolumn{2}{c||}{ {\rm energy estimates} {\rm }}\\
 \hline
  $d$ &{\rm   lower }
  &{\rm   upper }\\
 \hline \hline
 0&1 &1\\
 0.25&0.768972 &0.768974\\
 0.50&0.635528&0.635530\\
 0.75&0.590300&0.590301\\
 1.00&0.618910&0.618920\\
 1.50&0.801493&0.801494\\
 2.00&0.951410&0.951420\\
 $\infty$&1 &1\\
  \hline
 \hline \hline
\end{tabular}
\end{table}

In Table \ref{dowe} sampling the ground-state results, the
non-monotonicity of the $d-$dependence of the energy is slightly
counterintuitive but still easy to understand. It is obvious that
the growth of the positive shift $d>0$ leads, at the beginning, to
the broadening of the well without a significant change of its size
near the origin. This implies the possibility of an approximation of
the potential by a broader effective well, i.e., in a solvable
approximation, by $V^{(eff)} = \omega^2 x^2$ with $\omega<1$
yielding the smaller ground state energy. Later, i.e., at the larger
shifts $d$ the central repulsive spike starts growing more quickly,
separating the two HO wells and making them only weakly coupled.
This must lead,  approximately, to the double degeneracy of the
low-lying spectrum and, in particular, to the unavoidable return of
the ground state energy to its initial value at the sufficiently
large $d$s.

%

\begin{table}[h]
\caption{The shift-dependence of the 1st-excited-state energy.}
 \label{dowe1}
\centering
\begin{tabular}{||c||c|c||}
\hline \hline
  \multicolumn{1}{||c||}{\rm {\rm shift}}&
 \multicolumn{2}{c||}{ {\rm energy estimates} {\rm }}\\
 \hline
  $d$ &{\rm   lower }
  &{\rm   upper }\\
 \hline \hline
 0&3 &3\\
 0.25&2.483910 &2.483920\\
 0.50&2.060760&2.060770\\
 0.75&1.724710&1.724720\\
 1.00&1.468460&1.468470\\
 1.50&1.157479&1.157480\\
 2.00&1.035760&1.035770\\
 $\infty$& 1 &1\\
  \hline
 \hline \hline
\end{tabular}
\end{table}

Table \ref{dowe1} illustrates that as long as the wave function of
the first excited state vanishes in the origin, it is much less
sensitive to the influence of the central spike. According to our
tests, the energy just decreases with the growth of $d$ and, in
contrast to the behavior of the ground state energy, it never drops
below $E=1$ at the large shifts $d \gg 1$. Marginally we may add
that the test is facilitated by the observation that at the relevant
spectral bound $E=1$, the wave-function solutions with the correct
odd-parity behavior in the origin may be further simplified and
expressed in terms of error functions. This simplification of the
general solution facilitates also the demonstration of the
complex-value nature of the asymptotic nodal zero at any $E<1$.


%
%

\begin{table}[h]
\caption{Shift-dependence of the 2nd-excited-state energy (at $d=1$
the state is quasi-exact).}
 \label{dowe2}
\centering
\begin{tabular}{||c||c|c||}
\hline \hline
  \multicolumn{1}{||c||}{\rm {\rm shift}}&
 \multicolumn{2}{c||}{ {\rm energy estimates} {\rm }}\\
 \hline
  $d$ &{\rm   lower }
  &{\rm   upper }\\
 \hline \hline
 0&5 &5\\
 0.25&4.34600 &4.34700\\
 0.50&3.79410&3.79420\\
 0.75&3.34470&3.34471\\
 1&3&3\\
 1.50&2.64860&2.64870\\
 2.00&2.73500&2.73510\\
 $\infty$& 3 &3\\
  \hline
 \hline \hline
\end{tabular}
\end{table}

Table \ref{dowe2} shows that the $d-$dependence of the second
excited level (i.e., its decrease followed by the increase) is
similar to its ground-state predecessor. Figure \ref{ffit}
complements these results by a typical sample of the first three
wave functions at $d=3/2$.

\section{Conclusions\label{summaryb}}

In our present paper we sought for a nontrivial support of the
existing tendencies towards an extension of the class
of ``solvable'' 1D potentials $V(x)$ from completely analytic to
non-analytic at some points and, in particular, in the origin. We
believe that these tendencies are natural and well motivated.

A strengthening and/or independent complement of our present
arguments may be found in Ref.~\cite{[297]} where we sampled the
methodical as well as practical gains of the approach by introducing
another spiked, centrally symmetrized 1-D potential of the
Morse-oscillator type. In spite of its non-analyticity in the origin
the model was still shown to exhibit several features of the more
conventional complete exact solvability.

For the generic 1-D interactions characterized by the various
incomplete forms of solvability (like QES) the situation appears
more complicated because the presence of the non-analyticities at
some coordinates may restrict, severely, the practical as well as
methodical applicability of the incomplete constructions in quantum
physics. One can feel encouraged by the existence of
parallels between the one-dimensional and three dimensional central
interactions. The phenomenological appeal of the latter
models is usually perceived as independent of the presence or
absence of a non-analyticity in the origin.  {\it
Pars pro toto}, let us just mention the singular nature of the so
called shape-invariant realizations of supersymmetry in quantum
mechanics \cite{Cooper}.

There are no reasons for an asymmetric
treatment of solvability (and, in particular, of the QES models) in
one and three dimensions. As long as in 3-D models the singularities
in the origin are currently tolerated, a transfer of the {\em same}
``freedom of non-analyticity'' to 1-D systems is desirable. In our
present paper we showed that such a change of paradigm is also
productive (new models may be constructed) and satisfactory (in the
QES setting, the classification of models may be formulated as
starting from the most elementary quadratic polynomials).


After a removal of the QES constraints,
the next-to-elementary, NES nature of our present bound-state
problem (\ref{SEx}) + (\ref{dolenhol}) can be perceived, on a purely
methodical level, as an inspiration of possible piecewise-analytic
shifts, matchings and generalizations of many other traditional
phenomenological interactions. Even if we restrict
attention just to the present example (\ref{dolenhol}) offering a
tunable transition between a single- and double-minimum shape, we
would like to emphasize that it possesses several specific
advantages.

First of all, the exact analytic form of the related NES wave
functions offers an undeniable amendment, e.g., in comparison with
the conventional polynomial options as sampled by  the famous
Mexican-hat (MH) central potential
$$V^{(MH)}(|\vec{r}|)=-a|\vec{r}|^2+|\vec{r}|^4.$$ Thus, one of the
domains of possible applicability of our model (\ref{dolenhol})
might be sought in the advanced field theories of self-interacting
systems possessing the so called Nambu-Goldstone bosons in which the
use of benchmark $V^{(MH)}(|\vec{r}|)$ remains popular in spite of
its purely numerical nature (cf., e.g., picture Nr.~7 in
\cite{Goldstone}). Not insisting too much on the analyticity
of $V^{(MH)}(|\vec{r}|)$, the author of {\it loc. cit.} emphasized
that with $a>0$, the MH dynamical ansatz leads to the solutions
which ``have lower symmetry than the Lagrangian, and contain mass
zero bosons''.

We believe that in comparison with $V^{(MH)}(|\vec{r}|)$, our
present model $V^{}_{[d]}(x)$ could prove competitive, in spite of
its non-analyticity, not only in the above-mentioned area but also
in various more straightforward applications ranging from the
elementary quantum mechanics of atoms and molecules up to various
topics in condensed matter physics. In all of these areas the
possible phenomenological use of our NES quantum model (\ref{SEx}) +
(\ref{dolenhol}) might complement the {purely numerical} approaches
which are needed when the currently more popular polynomial
potentials are used.


\begin{thebibliography}{00}
\bibitem{[24a]}
R. Sasaki, Confining non-analytic exponential potential $V (x) = g^2
\exp(2|x|)$
and its exact Bessel-function solvability. ArXiv: 1611.02467.
\end{thebibliography}
\end{document}